# Localized states emerging from singular and nonsingular flat bands in a frustrated fractal-like photonic lattice

Haissam Hanafi, Philip Menz, and Cornelia Denz

*Institute of Applied Physics, University of Münster, Correnstr. 2, 48149 Münster, Germany*
*haissam.hanafi@uni-muenster.de*

**Abstract:** We report on singular and nonsingular flat bands in a Sierpinski fractal-like photonic lattice. We demonstrate that the the lowest two bands, being isolated and degenerate due to geometrical frustration, are nonsingular and thus can be spanned by a complete set of compact localized states. We experimentally prove these states to propagate diffractionless in the photonic lattice. Our results reveal the interplay between geometrical frustration, degenerate flat bands and compact localized states in a single photonic lattice, and pave the way to photonic spin liquid ground states.

Periodic systems, described by the Bloch formalism and governed by Schrödinger-like equations, are of importance for a huge variety of scientific fields. These systems possess a band structure that describes the energy levels of allowed and forbidden quantum mechanical states. One striking feature in such a band structure is the presence of completely flat bands that are dispersionless over the whole Brillouin zone. The energy of the states occupying such a band is independent of their momentum making it an ideal testbed to explore strong correlation physics as e.g. represented by the fractional quantum Hall effect [1] or Wigner crystallization [2]. In the context of the Bloch formalism, the flat band eigenstates of the tight-binding Hamiltonian are degenerate having the same eigenvalue. Flat bands are usually described by real space eigenfunctions that are compact and localized, the so-called compact localized states (CLSs) [3]. Localization is called compact when it has a strictly vanishing amplitude outside a finite region of the lattice. The existence of such states originates from destructive interference of the wave function after the hopping process described by the Hamiltonian.
In 2019 Rhim et al. proposed a classification of flat bands into singular and nonsingular ones based on the presence or absence of immovable discontinuities in their Bloch wave function in momentum space [4]. This classification can also be interpreted as the incompleteness or completeness of the set of CLSs in spanning the whole flat band. For a nonsingular flat band, there is no immovable discontinuity of its Bloch wave function. A discontinuity is immovable when there is no local gauge choice which, by shifting the position of the singular point, makes the Bloch wave function continuous [4]. A nonsingular flat band can be isolated from the dispersive bands, and the CLSs form a complete set. If, on the other hand, the band is singular, it possesses an immovable discontinuity which originates from the touching with at least one other dispersive band at the singular point. In this case, the CLSs turn out to be incomplete i.e. missing states which manifest nontrivial real-space topology must be complemented to form a complete set. These states are known as noncontractible loop states (NLSs), which are localized and compact in one direction, while they extend infinitely in the other direction and cannot be contracted by adding CLSs [5]. Up to now, singular and nonsingular flat bands have been discussed mainly in separate, tailored photonic lattices that exhibit explicitly the state to be investigated. For example, CLSs and NLSs of singular flat bands have been shown in the Lieb lattice [6–8], or the kagome lattice [9, 10], and CLSs of the nonsingular type in 1.5D lattices [11] or in a driven graphene lattice [12]). Only recently first efforts were made to combine the two classes of flat bands in an optically induced lattice in a photorefractive crystal [13]. However, up to now, no experimental realization proved the coexistence of this two topological classes of singular and nonsingular flat bands in a single lattice by analyzing the completeness conditions of the sets of CLSs. Here, we propose and experimentally demonstrate, for the first time to our knowledge, a fractal-like lattice that exhibits multiple flat bands fabricated by femtosecond direct laser writing on a large-scale fused silica chip. Crucially, we deliver the fundamental proof of singularity or nonsingularity of these bands by deriving and experimentally demonstrating their relative incomplete and complete sets of CLSs. The underlying fractal-like lattice was first described theoretically in 2018 by Pal et al. [14] and is constructed taking a first-generation Sierpinski gasket as the unit cell (Fig. 1 (a)). Note that, its band structure hosts two degenerate flat bands at lowest energy which we reveal to be nonsingular, as well as a singular flat band (Fig. 1 (d)).

The double degeneracy of the isolated flat bands in the Sierpinski fractal-like lattice is a feature this lattice shares with the theoretical kagome-3 model. This model was proposed in early studies on singular flat bands [5, 15] emphasizing the frustrated configuration that is inherent to the kagome lattice in general [16]. Geometrical frustration is best known in antiferromagnetic interactions when the lattice geometry inhibits the formation of a simple, ordered, spin configuration [17]. The resulting degenerate manifold of ground states, can be interpreted as a spin liquid. In our photonic realization of such geometrical frustration, we demonstrate that the degenerate flat bands of the Sierpinski fractal-like lattice, similar to those in the kagome-3 model, are nonsingular, while the zero energy flat band is singular. Additionally, we explain how the geometrical frustration of the lattice is linked to multiple degenerate flat bands.

For the experimental realization we chose a photonic system based on an array of evanescently coupled waveguides which is known as a photonic lattice [18, 19]. Photonic lattices represent an extremely effective platform for studying many intriguing phenomena related to periodic structures [20–23]. They have the advantage to allow accessing the flat band signature as light localization. The light propagation in such a lattice is governed by a Schrödinger-like paraxial wave equation, with the band structure representing a spatial diffraction relation [24]. Compared to an atomic lattice where the transport dynamics of electrons are observed, a photonic lattice is characterized by the light field evolution in propagation direction describing photon transport. The role of the energy in the dispersion relation is taken by the propagation constant $\beta$.

In a photonic lattice, the realization of the kagome-3 model is not trivial since it assumes equal coupling strength for nearest and next-nearest neighbours, whereas in a photonic lattice the coupling is proportional to the waveguide separation. However, the Sierpinski fractal-like lattice shows doubly degenerate flat bands as the kagome-3 model does and includes even richer features. The Sierpinski fractal-like lattice can be obtained by a 'decoration' of the kagome lattice with additional lattice sites. Decorating lattices is a well known technique in the context of localization and flat bands [25, 26]. Adding three lattice sites to the triangular kagome unit cell in between existing ones, and spacing them uniformly to obtain equal coupling, leads to the Sierpinski gasket unit cell shown in Fig. 1 (a). In order to calculate the band structure of the lattice we chose a tight-binding approximation [27]. We assume equal hopping amplitude between nearest neighbors, and set it to 1 (together with the lattice constant $d = 1$) for simplicity. We then obtain a 6x6 Hamiltonian in momentum space whose dimension reflects the size of the unit cell (see Supplementary Material for detailed calculation). By solving the eigenvalue problem, we obtain the band structure shown in Fig. 1 (d) consisting of six bands. The lowest two of them, located at $\beta = -2$, are degenerated, isolated from the dispersive bands, and completely flat over the whole Brillouin zone. There is a third flat band at $\beta = 0$ that touches two other dispersive bands at $\mathbf{k} = 0$ forming a spin-1 Dirac cone like in the kagome with staggered flux [28], or in the Lieb lattice [29]. To obtain the CLSs of the flat bands, we perform an inverse Fourier transform of the corresponding eigenvectors of the Hamiltonian [4] (see Supplementary Material for detailed calculation). To validate the found CLSs, we prove that they fulfil a Schrödinger equation of the form

$$(\beta - \epsilon_i)\psi_i = \sum_j \tau_{ij}\psi_j, \qquad (1)$$

where $\epsilon_i$ is the on-site potential at the $i$th site, $\tau_{ij}$ is the hopping amplitude between neighboring sites, and $\psi_i$ is the wave function amplitude at the $i$th site [14]. In our case we assume all sites to have equal potential and shift this value to zero and consider only nearest neighbors with a hopping amplitude $\tau_{ij} = 1$.

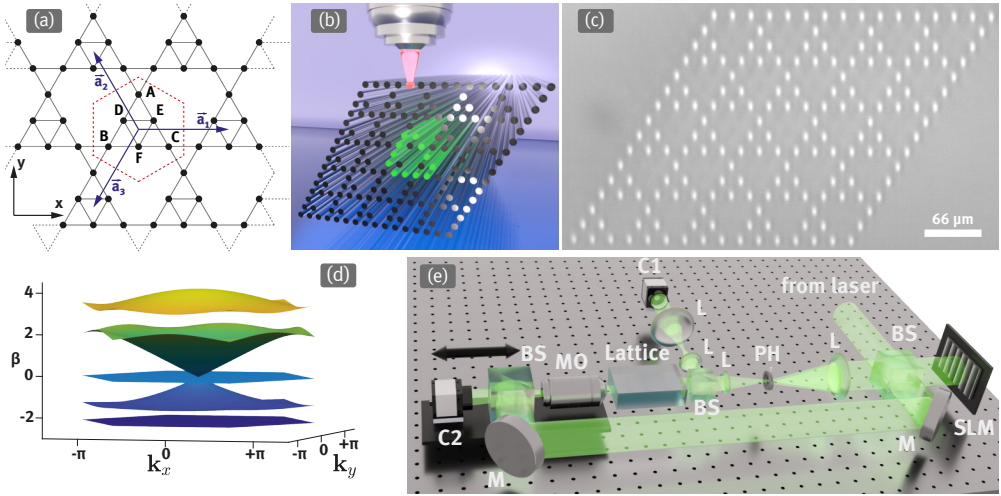

Fig. 1. (a) Schematic of the Sierpinski lattice. Black dots represent lattice sites; grey lines connecting the lattice site indicate NN hopping of equal strength; the red dashed hexagon marks a unit cell composed of 6 lattices sites numbered from A to F; $\mathbf{a}_{1,2,3}$ are lattice vectors connecting neighboring unit cells. (b) Femtosecond direct laser writing technique. (c) White light microscope image of the lattice in the crystal. (d) Nearest neighbor tight-binding band structure of the Sierpinski lattice in the first Brillouin zone. (e) Experimental system for probing the flat band. BS: beam splitter; SLM: phase-only spatial light modulator, M: mirror; L: lens; MO: microscope objective; PH: pinhole; C1/C2 camera 1/2.

The CLS from the flat band at $\beta = 0$ is shown in Fig. 2 (a1). In agreement with [14], it takes the form of a truncated triangle and corresponds to an extension of the normal kagome CLS. Due to the decoration of the kagome unit cell, the flat band is shifted from $\beta = -2$ to $\beta = 0$ [26]. Like in the kagome lattice, $N$ translated copies of the CLS are linearly dependent as a result of the discontinuity of the eigenvector at the centre of the Brillouin zone. The discontinuity is immovable, i.e. the band touching is singular. Therefore, the flat band cannot be completely described by the CLSs, and some NLSs exist. The NLSs of the Sierpinski fractal-like lattice can be derived considering NLSs of the normal kagome lattice. In the kagome lattice, the NLSs consist of lines of lattice sites with the same amplitude but alternating opposite phase on the lattice axes [5]. In the Sierpinski lattice, they essentially remain the same, only being decorated (see Supplementary Material). Like in the kagome lattice, as they extend infinitely in one direction, the NLSs are stable only under periodic boundary conditions. In systems with finite open boundary conditions, they cannot be obtained (unlike in the case of the Lieb lattice where designed boundaries allow this [8]). However, to realize the NLSs, the lattice has either to be constructed with a specific periodic boundary geometry (Corbino geometry), or the mode has to be spanned on a closed contour along the boundary of a finite lattice realizing a robust boundary mode (RBM) [10].

Considering the flat bands at $\beta = -2$, at first sight, the situation seems to be similar. The two eigenvectors are discontinuous at two distinct points in the Brillouin zone (see Supplementary Material for calculations). This means that the CLSs we obtain by inverse Fourier transform do not form a complete set, i.e. they cannot span the flat band completely (note that in [14] for $\beta = -2$ only one CLS was proposed, it has a less compact form, and also does not form a complete set). Thus, additional NLSs are needed which intuitively can be identified being similar to the $\beta = 0$ ones, but with alternating opposite phases (see Supplementary Material). However, the two flat bands at $\beta = -2$ are isolated from the dispersive bands, i.e. there is no band touching (besides from the complete degeneracy itself). Therefore, these bands are nonsingular, and thus a complete set of CLSs can be found. In fact, it turns out that, exactly as in the kagome-3 model, it is possible to find two nonsingular eigenvectors and their related CLSs which make the supposed NLSs contractible. The key point is that the flat bands are completely degenerate. It is hence possible to mix the two eigenvectors by linearly combining them in order to obtain nonsingular ones which do not have discontinuities. As before, from this new eigenvectors, we obtain an expression for the CLSs (see Supplementary Material). These CLSs are shown in Fig. 2 (b1) and (c1), respectively. Due to their appearance we name this states *reindeer* CLS-1 and *reindeer* CLS-2. Since this set of CLSs is complete, it is possible to construct the supposed NLSs as well as any other state on the flat band by a linear combination of them.

Most of the early work on singular flat bands and NLSs focused on lattices exhibiting strong geometrical frustration [5]. It was however demonstrated that this is not a necessary condition [8, 30]. The NLSs are a direct manifestation of the immovable discontinuities of the Bloch functions generated by the band touching [4]. Extending this findings, we demonstrate that flat bands of a geometrically frustrated lattice can be singular and nonsingular. Note that in the fractal-like lattice there is a relation between degenerate flat bands and geometrical frustration. This relation can be seen in the form of the *reindeer* CLSs: if we try to construct a CLS into the first-generation Sierpinski triangle unit cell, we cannot simultaneously obtain destructive interference in all three corners of the triangle (lattice sites A,B and C), due to this optical analog to geometrical frustration. If we choose destructive interference in B and C, we end up with the *reindeer* CLS-1 as smallest possible CLS; if instead we want to achieve destructive interference in A and B, we get the *reindeer* CLS-2. Additionally, if instead of the first generation, we choose a second-generation Sierpinski triangle as the unit cell, we will obtain five completely degenerate flat bands [14]. Thus, we can conclude that frustrated geometry as well as fractal self similarity leads to multiple degenerate flat bands.

We experimentally realize the previously discussed localized flat band states of the Sierpinski fractal-like lattice by fabricating a sample composed of 181 single-mode waveguides arranged in 5x5 unit cells with a rhombic termination. The single-mode waveguides are induced by femtosecond direct laser writing in a fused silica glass chip (Ultrapure Synthetic Fused Silica SQ0) [31–33]. The induction laser emits light at a central wavelength of 1030 nm, with a pulse duration of 250 fs at a repetition rate of 125 kHz. In order to ensure circular and homogeneous single-mode waveguides over the entire extent of the lattice, we apply a slit beam shaping method in combination with a spherical aberration correction [34]. The induction laser beam is hereby shaped by a single phase-only spatial light modulator (SLM). A white light microscope image of the front-facet of the fabricated lattice is shown in Fig. 1 (b). A lattice constant of $d = 66\,\mu$m (waveguide separation of $22\,\mu$m) is chosen. This ensures sufficient coupling during z-propagation in the 2 cm sample to distinguish clearly between diffracting and nondiffracting states, while also being large enough to prevent next-nearest neighbor coupling disturbing the flatness of the bands. We excite the flat band states by using the experimental system depicted in Fig. 1 (e). The probing continuous wave (cw) laser beam operates at a wavelength of 532 nm, and is modulated in amplitude and phase with a phase-only SLM to generate the different CLS-light fields [35]. To ensure high coupling of the CLS-light fields into the photonic lattice, we monitor the front-facet of the lattice and align in to the back-reflection. We record the output by imaging the lattice back-facet onto a camera using a microscope objective. The phase of the light field is recorded by interference with a tilted plane wave and reconstructed via a digital holographic method [36].

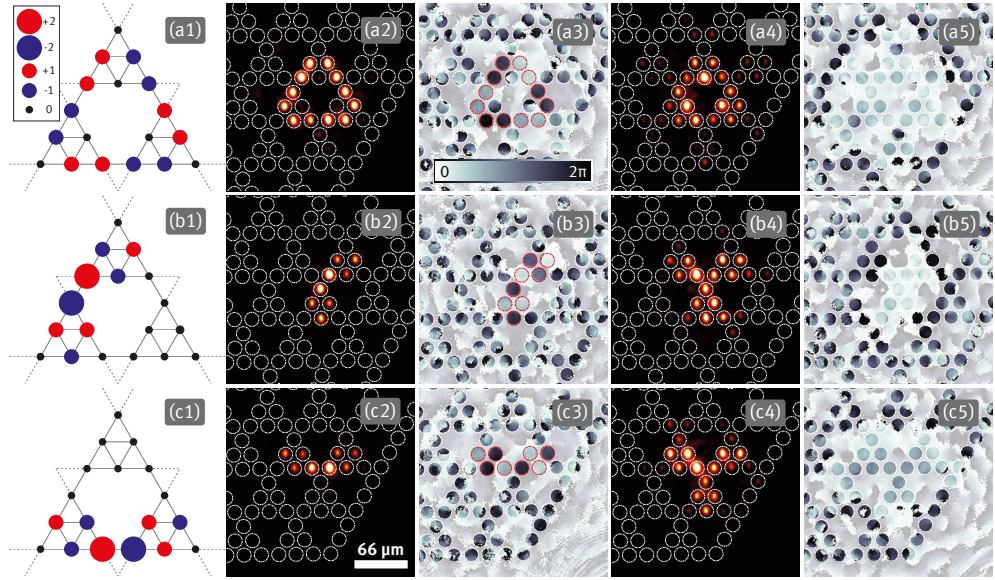

Fig. 2. CLSs and experimental demonstration of their diffraction-free propagation in the Sierpinski fractal-like lattice. (a1) CLS at $\beta = 0$; intensity (a2) and phase (a3) of the CLS light field at $\beta = 0$ after propagation in the lattice; intensity (a4) and phase (a5) of a light field with the same amplitude of the CLS at $\beta = 0$ but with flat phase after propagation in the lattice. (b1) *reindeer* CLS-1; (b2)-(b5) same as (a2)-(a5) but for the *reindeer* CLS-1. (c1) *reindeer* CLS-2; (c2)-(c5) same as (a2)-(a5) but for the *reindeer* CLS-2.

We now experimentally demonstrate flat bands of singular and nonsingular nature in a single photonic lattice. First, we investigate the CLS of the flat band located at $\beta = 0$. The light field exciting the CLS is composed of Gaussian spots of alternating opposite phase arranged in a truncated triangle and grouped two-by-two (Fig.2 (a1)). We compare the propagation of this light field in the photonic lattice with a light field with the same intensity distribution but equal phase at all excited lattice sites. This state will excite not only the flat band, but also diffracting ones. The outcome should therefore be clearly distinguishable from the CLS one. The propagated lights fields, which we measure at the back-facet of the lattice, are depicted in Fig.2 (a2)-(a5). In Fig.2 (a2) and (a3) we observe the light field exciting the CLS propagating without diffraction, in a robust and localized way over the whole lattice length of 2 cm. In contrast, as shown in Fig.2 (a4) and (a5), a light field having a flat phase strongly diffracts, and the initial light field distribution can no longer be recognized in the output. The same clear results are obtained for the two degenerated flat bands at $\beta = -2$ and the corresponding *reindeer* CLSs. The light fields corresponding to the CLSs of the degenerate flat bands (Fig.2 (b1) and (c1)) propagate diffraction free, and the initial state is preserved (Fig.2 (b2)-(b3) and (c2)-(c3)). On the other hand, for a light field with the same intensity distribution but equal phase at all excited lattice sites, we observe strong diffraction into the photonic lattice (Fig.2 (b4)-(b5) and (c4)-(c5)).

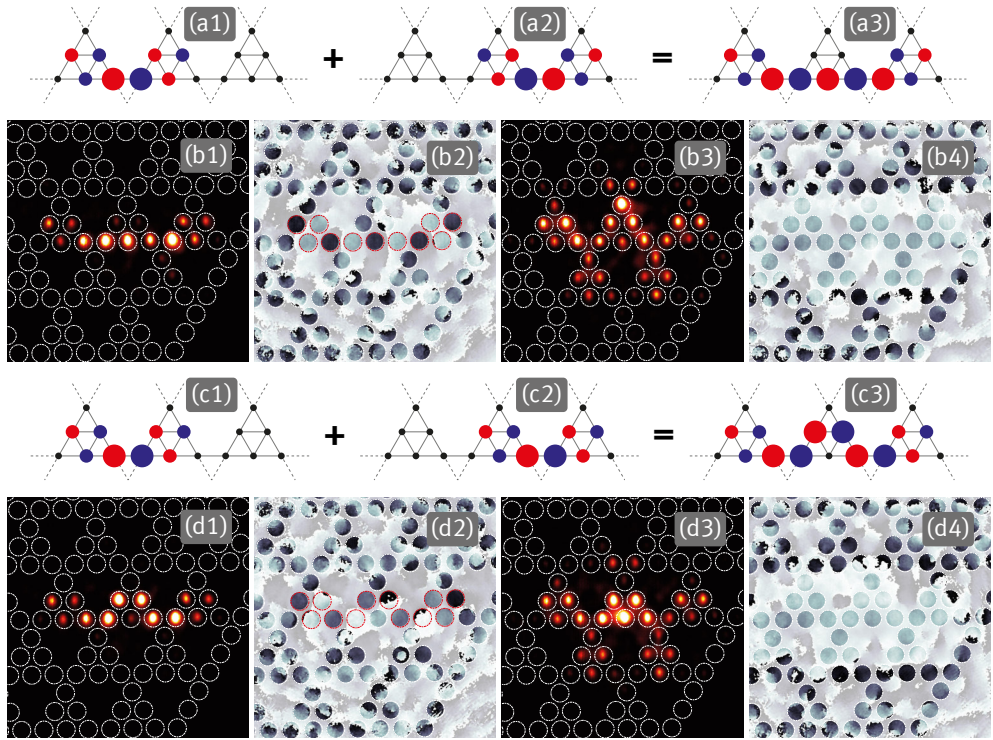

Fig. 3. Experimental demonstration of diffraction-free propagation for linear combinations of *reindeer* CLSs in the Sierpinski fractal-like lattice. (a1) *reindeer* CLS-2. (a2) $\pi$-shifted *reindeer* CLS-2 translated one unit cell to the right. (a3) linear combination of (a1) and (a2). Intensity (b1) and phase (b2) of the light field corresponding to (a3) propagated in the lattice. (b3)-(b4) same as (b1)-(b2) but with equal phase at all excited lattice sites as input. (c1) *reindeer* CLS-2. (c2) *reindeer* CLS-2 translated one unit cell to the right. (c3) linear combination of (c1) and (c2). (d1)-(d2) same as (b1)-(b2) but for (c3) as input. (d3)-(d4) same as (d1)-(d2) but with equal phase at all excited lattice sites as input.

Thus, the *reindeer* CLSs are a complete set spanning the whole flat band. To demonstrate this, we prove that the supposed NLSs are contractible, as they can be constructed by a linear superposition of *reindeer* CLSs. In Fig. 3, we show a linear superposition of two horizontally-shifted *reindeer* CLSs that are either out of phase (a3), or in phase (c3). By adding more *reindeer* CLSs, the zigzag or straight lines can be extended indefinitely. We demonstrate that these line segments remain intact during propagation in the lattice (Fig. 3 (b1)-(b2) and (d1)-(d2)), while the line segments with flat phase diffract during propagation (Fig. 3 (b3)-(b4) and (d3)-(d4)). Though within this paper we only demonstrate linear superposition of one type of *reindeer* CLS, the findings hold true also for the other *reindeer* CLSs, and the associated line segments (see Supplementary Material).

In conclusion, we have demonstrated singular and nonsingular flat bands in a first-generation Sierpinski fractal-like photonic lattice fabricated by femtosecond direct laser writing on a large-scale fused silica chip. We explained that the Sierpinski fractal-like lattice is constructed by a decoration of the kagome lattice and has three flat bands. We revealed that one flat band is singular as in the kagome lattice, while the other two are degenerate, isolated and therefore nonsingular as in the kagome-3 model [5, 15]. We theoretically derived the compact localized states of the singular and nonsingular flat bands, and experimentally proved them to propagate diffractionless in the photonic lattice. We confirm the nonsingularity of the flat bands by showing theoretically and experimentally that with the right choice of CLSs, the *reindeer*-CLSs, every supposed NLS is a superposition of them. Our results validate the classification of flat bands according to the band-crossing singularity of the respective Bloch wave function [4]. Our findings allow a general insight into the nature of degenerate flat band systems, and pave the way to photonic lattice realizations based on frustrated interactions, including photonic analogues of spin liquid phenomena.